\documentclass[aip,pop,reprint,amsmath,amssymb]{revtex4-2}
\usepackage[T1]{fontenc}
\usepackage{bm}
\usepackage{graphicx}
\usepackage[colorlinks=true,citecolor=blue,linkcolor=blue,urlcolor=blue]{hyperref}

\newcommand{\PT}{\mathrm{PT}}
\newcommand{\HVM}{\mathcal{H}_{\mathrm{VM}}}
\newcommand{\HVP}{\mathcal{H}_{\mathrm{VP}}}
\newtheorem{theorem}{Theorem}

\begin{document}
\title{Parity-Time Symmetry of Plasma Waves}
\author{Hong Qin}
\affiliation{Department of Nuclear Engineering and Engineering Physics, University of Wisconsin-Madison, Madison, WI 53706, USA}

\begin{abstract}
The susceptibility tensor of a homogeneous collisionless magnetized plasma possesses parity-time (PT) symmetry. For a stationary gyrotropic equilibrium, with the wave vector $\mathbf{k}$ and the equilibrium magnetic field $\mathbf{B}_0$ in the $x$--$z$ plane, the kinetic response satisfies $\chi(\omega^*,\mathbf{k})=R\chi(\omega,\mathbf{k})^*R$, where $R=\operatorname{diag}(1,-1,1)$. We show that the susceptibility PT-symmetry is a reduced-response-space manifestation of PT symmetry of the underpinning kinetic system. We identify parity $P$ and time reversal $T$ on the complete electromagnetic kinetic eigenmode space and show that the Hamiltonian operator $\HVM$ of the linearized Vlasov--Maxwell system is PT-symmetric, satisfying $PT\HVM(PT)^{-1}=\HVM$.  The corresponding PT symmetry of the homogeneous magnetized Vlasov--Poisson eigenproblem is also established. Many kinetic instabilities associated with nonmonotonic distributions, rings, loss cones, and resonant energetic-particle populations are fundamentally spontaneous PT-symmetry breaking, which is achieved through and only through Krein collisions. Three examples demonstrate distinct utilities of this analysis. Definite-sign cyclotron residues protect thermal Bernstein modes from spontaneous PT-symmetry breaking. The nonmonotonic Dory--Guest--Harris distribution permits sign-indefinite residues and Krein collisions, for which we derive an analytical instability threshold and growth rate. The recently discovered mode-pole resonance instability provides a distinct realization of spontaneous PT-symmetry breaking and illustrates how PT symmetry can guide the discovery of kinetic instabilities with applications to fusion energy. These results establish PT symmetry as a structural principle of plasma waves and instabilities and open new directions for studying spectral topology in kinetic plasmas.  
\end{abstract}
\maketitle

\section{Introduction}
\label{sec:introduction}

\begin{figure}[tb]
 \centering
 \includegraphics[width=\columnwidth]{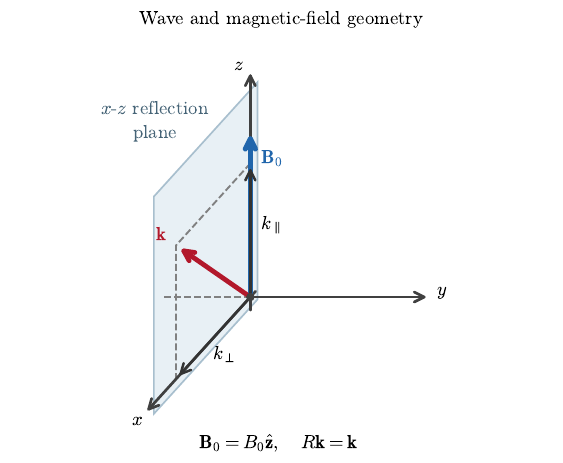}
 \caption{Geometry of electromagnetic waves in a magnetized plasma and the reflection $R=\operatorname{diag}(1,-1,1)$. The equilibrium magnetic field lies along $z$, and $\mathbf{k}=k_\perp\hat{\mathbf{x}}+k_\parallel\hat{\mathbf{z}}$ lies in the shaded $x$--$z$ reflection plane.  $R$ reverses the $y$ component while preserving the $x$ and $z$ components.}
 \label{fig:geometry-reflection}
\end{figure}

The kinetic plasma susceptibility tensor is one of the most familiar objects in plasma waves \cite{Stix1992,Dodin,Qin2007}. It describes the electromagnetic response after the perturbed distribution functions have been eliminated from the linearized Vlasov--Maxwell equations. We begin with a symmetry of this standard kinetic response. Consider a homogeneous collisionless plasma with a stationary gyrotropic equilibrium and choose the geometry
\begin{equation}
 \mathbf{B}_0=B_0\hat{\mathbf{z}},\qquad
 \mathbf{k}=k_\perp\hat{\mathbf{x}}+k_\parallel\hat{\mathbf{z}},
 \label{eq:geometry}
\end{equation}
where $\mathbf{k}$ is real. The $x$--$z$ plane contains both $\mathbf{k}$ and $\mathbf{B}_0$. Reflection across this plane is represented by
\begin{equation}
 R=\operatorname{diag}(1,-1,1).
 \label{eq:reflection}
\end{equation}
$\mathbf{k}$ is invariant under this reflection, i.e., $R\mathbf{k}=\mathbf{k}$.  Figure~\ref{fig:geometry-reflection} illustrates this geometry and the reflection plane of $R$. The formal hot-plasma susceptibility satisfies
\begin{equation}
 \chi(\omega^*,\mathbf{k})=R\chi(\omega,\mathbf{k})^*R,
 \label{eq:chi-pt}
\end{equation}
with $\mathbf{k}$ and $\mathbf{B}_0$ held fixed. Here and throughout, the asterisk denotes complex conjugation. At the electromagnetic response level, $R$ represents parity on the electric-field subspace, while complex conjugation represents time reversal in the spectral formulation. We therefore refer to Eq.~\eqref{eq:chi-pt} as the PT symmetry of the magnetized Vlasov--Maxwell susceptibility tensor.

This identity immediately implies the corresponding symmetry of the electromagnetic dispersion matrix $M(\omega,\mathbf{k})$ and its determinant,
\begin{align}
 M(\omega^*,\mathbf{k})&=RM(\omega,\mathbf{k})^*R,\\
 \det M(\omega^*,\mathbf{k})&=[\det M(\omega,\mathbf{k})]^*.
 \label{eq:dispersion-pt}
\end{align}
Consequently, the formal electromagnetic eigenfrequencies occur in conjugate pairs: if $\omega$ is an eigenfrequency, then $\omega^*$ is also an eigenfrequency. The algebraic identity can be verified directly in two independent representations of the susceptibility. The conventional representation establishes it harmonic by harmonic in the infinite cyclotron-harmonic sum. The representation of Qin, Phillips, and Davidson~\cite{Qin2007} establishes the same identity using Bessel functions of complex order without an infinite harmonic sum. The appearance of the same symmetry in both representations suggests an origin in the underlying kinetic dynamics.

The central question of this paper is why the standard magnetized Vlasov--Maxwell susceptibility possesses this PT symmetry and what this symmetry implies for kinetic stability. The susceptibility is obtained by solving the linearized Vlasov equation for the perturbed distribution function and substituting the resulting current into Maxwell's equations. Its symmetry may therefore be inherited from the complete kinetic system before these degrees of freedom are eliminated. We show that this is indeed the case. After identifying parity $P$ and time reversal $T$ on the complete  electromagnetic kinetic eigenmode space, we establish
\begin{equation}
PT\HVM(PT)^{-1}=\HVM, 
 \label{eq:vm-pt}
\end{equation}
where $\HVM$ is the unreduced linearized Vlasov--Maxwell Hamiltonian operator. 

\section{Overview and main results}

\subsection{PT symmetry}
\label{sec:pt-overview}

For a non-Hermitian Hamiltonian, the absence of Hermiticity invites consideration of other symmetries that can constrain its physics. Parity-time symmetry has a spacetime motivation in the Lorentz group $O(1,3)$, the group of homogeneous transformations preserving the Minkowski metric. This group has four disconnected components, with the identity component given by the proper orthochronous subgroup $SO^+(1,3)$. Invariance under this connected subgroup does not require invariance under the remaining components. The full group can be written as
\begin{equation}
 O(1,3)=SO^+(1,3)\rtimes
 \{I,\mathcal{P},\mathcal{T},\mathcal{P}\mathcal{T}\},
 \label{eq:lorentz-components}
\end{equation}
where $\mathcal{P}=\operatorname{diag}(1,-1,-1,-1)$ reverses space and $\mathcal{T}=\operatorname{diag}(-1,1,1,1)$ reverses time. Since parity and time reversal are not separately universal symmetries of physical interactions, a natural question is whether their product can remain a symmetry when either operation alone fails.\cite{Bender2019} Bender and Boettcher initiated the study of non-Hermitian Hamiltonians with this combined symmetry,\cite{Bender1998} and subsequent work developed its spectral theory, quantum dynamics, coupled-oscillator realizations, and field-theoretic extensions.\cite{Bender2007,Bender2010,Bender2007a,Bender2014,Felski2021} Broader accounts are given in the book by Bender and collaborators\cite{Bender2018} and the review by Bender and Hook.\cite{Bender2024}

The spacetime transformations motivate the symmetry, while its representations must be specified for each eigenproblem. In particular, the spectral parity $P$ used below represents reflection across the $\mathbf{k}$--$\mathbf{B}_0$ plane on the space of distribution functions and fields. 

Consider an eigenvalue problem $H\psi=\omega\psi$. Parity is represented by a linear involution $P$, satisfying $P^2=I$. Time reversal changes $i$ to $-i$; in the complex spectral representation used here it is represented by complex conjugation, $T=K$. PT symmetry requires
\begin{equation}
 PTH(PT)^{-1}=H \quad  \text{or} \quad  PH^{*}P^{-1}=H, 
 \label{eq:general-pt}
\end{equation}
Neither parity nor time reversal needs to leave $H$ invariant separately. Their successive action is the symmetry of the eigenproblem.

For $H\psi=\omega\psi$, Eq.~\eqref{eq:general-pt} gives
\begin{equation}
 H(\PT\psi)=\omega^*(\PT\psi).
 \label{eq:partner}
\end{equation}
The spectrum is therefore symmetric with respect to the real axis. When an eigenmode preserves PT symmetry, its transformed state is proportional to itself,
\begin{equation}
 \PT\psi=\lambda\psi,
 \label{eq:preserving}
\end{equation}
where $\lambda$ is a nonzero scalar. Equations~\eqref{eq:partner} and \eqref{eq:preserving} then require $\omega=\omega^*$, so the eigenfrequency is real. 

A simple real-frequency eigenmode is preserved by PT up to a phase. When a mode has a nonreal eigenfrequency, PT maps it into a distinct conjugate partner,
\begin{equation}
 \PT\psi_+\propto\psi_-,\qquad
 \omega_-=\omega_+^*.
 \label{eq:broken}
\end{equation}
For this broken-symmetry pair, the eigenfrequencies have the form
\begin{equation}
 \omega_\pm=\omega_r\pm i\gamma,\qquad \gamma>0.
 \label{eq:growing-decaying}
\end{equation}
Note that the governing operator remains PT symmetric. Spontaneous PT symmetry breaking refers to the fact that the symmetry of the individual eigenmodes is broken. 

\subsection{PT symmetry in plasma waves}
The original physical interpretation of PT symmetry is balanced loss and gain in coupled quantum subsystems \cite{Bender2014}.  We recently established that conservative classical systems, including neutral fluids and fluid plasmas, admit PT symmetry\cite{Qin2019KH,Zhang2020PT,Fu2020KH,Qin2021}, although these conservative systems require no externally imposed gain or loss. More importantly,  in these conservative classical systems, spontaneous PT symmetry breaking leads to instabilities, including the Kelvin--Helmholtz instability\cite{Qin2019KH,Fu2020KH} and drift-wave instability \cite {Qin2021}. 
With the normal-mode convention $\exp(i\mathbf{k}\cdot\mathbf{x}-i\omega t)$, the two modes in the complex-conjugate pair in Eq.~\eqref{eq:growing-decaying} grow and decay, respectively. 

It is also proven that PT symmetry implies pseudo-Hermiticity in finite-dimensional systems without assuming diagonalizability\cite{Zhang2020PT}.  Therefore, spontaneous PT symmetry can only be achieved through the Krein collision of a pair of real eigenmodes with opposite signatures. These results established PT symmetry as a useful framework for instabilities in classical fluids and plasmas \cite{Qin2019KH,Zhang2020PT,Fu2020KH,Qin2021,Qin2026TS,Qin2026IBW,Meng2023}. 

\subsection{Main results}
The present work extends this program from the spectra of neutral fluids and fluid plasmas to the magnetized Vlasov--Maxwell and Vlasov--Poisson eigenproblems. 

The logical hierarchy is that PT symmetry of $\HVM$ implies PT symmetry of $\chi$, which in turn implies conjugation symmetry of the electromagnetic spectrum. In particular, the susceptibility symmetry can be deduced from the unreduced kinetic equations without explicitly solving for the perturbed distribution functions. We also establish the corresponding PT symmetry of the homogeneous magnetized Vlasov--Poisson system. 

The distinction between the susceptibility and the unreduced eigenproblem is essential for the physical analysis. Equation~\eqref{eq:chi-pt} constrains the electromagnetic dispersion spectrum, whereas the complete eigenvector contains both the perturbed distribution functions and the electromagnetic fields. Identifying $P$ and $T$ on this state space informs whether an individual kinetic eigenmode is preserved under the PT transformation. Many kinetic instabilities in magnetized collisionless plasmas, particularly those associated with nonmonotonic distributions, rings, loss cones, and resonant energetic-particle populations, can be viewed as spontaneous PT-symmetry breaking. This suggests a common framework for understanding how velocity-space structure permits unstable kinetic modes.

We study three examples to demonstrate the important utilities of PT-symmetry analysis. For thermal Bernstein modes at strictly perpendicular propagation, definite-sign cyclotron residues enforce a monotonic dispersion function between neighboring poles and protect the real branches from the Krein collision required for spontaneous PT breaking. The symmetry analysis therefore explains why instability is prohibited in this system. For the Dory--Guest--Harris (DGH) distribution, nonmonotonic dependence on perpendicular velocity permits cyclotron residues of opposite signs. This removes the monotonicity protection and allows real modes to collide and form a growing and decaying pair. We derive an analytical double-root condition for the onset of this transition and identify the necessary residue-sign condition for the collision. PT symmetry analysis thus yields a new analytical description of the DGH instability. In the third example, resonance between a collective plasma mode and a kinetic pole provides a distinct mechanism for spontaneous PT breaking. PT-symmetry analysis guided the discovery of this mode-pole instability and supplies an analytical criterion for its onset. Together, these examples illustrate the explanatory, analytical, and predictive uses of kinetic PT symmetry analysis. The explicit kinetic eigenoperator also provides a foundation for future investigations of topology and spectral flow in PT-symmetric kinetic plasmas.

The paper is organized as follows. Section~\ref{sec:susceptibility} establishes the PT symmetry of the hot-plasma susceptibility tensor and the associated electromagnetic dispersion relation using two complementary susceptibility representations. Its final subsection, Sec.~\ref{sec:kinetic-origin}, examines the origin of this reduced response symmetry and traces it to the unreduced kinetic eigenproblem. In Section~\ref{sec:vlasov-maxwell}, parity and time reversal are defined for the full linearized Vlasov--Maxwell system, and invariance under their successive action is proved. Section~\ref{sec:vlasov-poisson} turns to the magnetized Vlasov--Poisson system and establishes the corresponding PT symmetry and conjugate spectral structure. The physical consequences are developed in Section~\ref{sec:physical-consequences}, including spontaneous PT-symmetry breaking, Krein collision, and applications to thermal Bernstein modes, DGH instability, and mode-pole instability. Section~\ref{sec:discussion} summarizes the main results and discusses the distinction between spectral and nonlinear time reversal and between the formal kinetic spectrum and Landau's causal response. Section~\ref{sec:future-work} presents directions for kinetic topology and explicit PT-symmetry breaking, with potential applications to fusion energy research.

\section{PT Symmetry of the Magnetized Vlasov--Maxwell Susceptibility Tensor}
\label{sec:susceptibility}

Consider an electromagnetic wave in the homogeneous collisionless magnetized plasma in the geometry defined by Eqs.~\eqref{eq:geometry} and \eqref{eq:reflection}. The equilibrium magnetic field is $\mathbf{B}_0=B_0\hat{\mathbf{z}}$, and the real wave vector is $\mathbf{k}=k_\perp\hat{\mathbf{x}}+k_\parallel\hat{\mathbf{z}}$. The reflection $R=\operatorname{diag}(1,-1,1)$ leaves $\mathbf{k}$ invariant and represents parity on the electric-field subspace. We show that the familiar hot-plasma susceptibility possesses PT symmetry.

\begin{theorem}
\label{thm:susceptibility}
\textup{(PT symmetry of Vlasov--Maxwell susceptibility)}
For the  linear electromagnetic response of a homogeneous collisionless magnetized plasma with a stationary gyrotropic equilibrium, the susceptibility tensor satisfies
\begin{equation}
 \chi(\omega^*,\mathbf{k})=R\chi(\omega,\mathbf{k})^*R.
 \label{eq:response-theorem}
\end{equation}
The electromagnetic dispersion matrix consequently satisfies
\begin{equation}
 M(\omega^*,\mathbf{k})=RM(\omega,\mathbf{k})^*R.
 \label{eq:matrix-theorem}
\end{equation}
Thus, the eigenfrequencies occur in complex-conjugate pairs.
\end{theorem}

Here the susceptibility is the formal response evaluated on the undeformed real velocity contour, at complex frequencies where the integrals and sums converge. The distinction from the analytically continued causal response is discussed in Sec.~\ref{sec:landau-discussion}.

We prove the susceptibility identity \eqref{eq:response-theorem} in two independent representations and then establish its consequences for the electromagnetic dispersion relation.

\subsection{Proof from the conventional cyclotron-harmonic representation}

For species $s$, let $f_{0s}=F_{0s}/n_{0s}$ be the equilibrium velocity distribution normalized to unity. Define the signed cyclotron frequency $\Omega_s=q_sB_0/m_s$ and the plasma frequency $\omega_{ps}^2=n_{0s}q_s^2/(\epsilon_0m_s)$. It is convenient to use the gyrotropic integration measure
\begin{equation}
 \int d\Gamma\equiv 2\pi\int_0^\infty v_\perp\,dv_\perp
 \int_{-\infty}^{\infty}dv_\parallel.
 \label{eq:velocity-measure}
\end{equation}
In this notation, the conventional susceptibility can be written as~\cite{Stix1992,Dodin,Qin2007}
\begin{align}
 \chi_s(\omega,\mathbf{k})={}&\chi_s^{\mathrm{nr}}(\omega)
 +\frac{\omega_{ps}^2}{\omega}\int d\Gamma\,v_\perp U_s
 \sum_{n=-\infty}^{\infty}\frac{\mathsf{T}_{sn}}{\Delta_{sn}},
 \label{eq:harmonic-response}\\
 \Delta_{sn}={}&\omega-k_\parallel v_\parallel-n\Omega_s,
 \label{eq:resonance-denominator}
\end{align}
where the nonresonant contribution is
\begin{equation}
 \chi_s^{\mathrm{nr}}(\omega)=\frac{\omega_{ps}^2}{\omega^2}
 \int d\Gamma\,v_\parallel
 \left(\frac{\partial f_{0s}}{\partial v_\parallel}
 -\frac{v_\parallel}{v_\perp}\frac{\partial f_{0s}}{\partial v_\perp}\right)
 \hat{\mathbf{z}}\hat{\mathbf{z}},
 \label{eq:nonresonant-response}
\end{equation}
and the scalar weight for the resonant term is
\begin{equation}
 U_s=\frac{\partial f_{0s}}{\partial v_\perp}
 +\frac{k_\parallel}{\omega}
 \left(v_\perp\frac{\partial f_{0s}}{\partial v_\parallel}
 -v_\parallel\frac{\partial f_{0s}}{\partial v_\perp}\right).
 \label{eq:scalar-weight}
\end{equation}
We assume sufficient regularity and decay of the equilibrium for the susceptibility integrals and harmonic sums to be well defined.

Let $\lambda=k_\perp v_\perp/\Omega_s$ and $r=v_\parallel/v_\perp$. With $J_n=J_n(\lambda)$ and $J_n'=dJ_n(\lambda)/d\lambda$, the cyclotron-harmonic tensor is
\begin{equation}
 \mathsf{T}_{sn}=\begin{pmatrix}
 \dfrac{n^2J_n^2}{\lambda^2}&
 \dfrac{inJ_nJ_n'}{\lambda}&
 \dfrac{nrJ_n^2}{\lambda}\\[5pt]
 -\dfrac{inJ_nJ_n'}{\lambda}&(J_n')^2&-irJ_nJ_n'\\[5pt]
 \dfrac{nrJ_n^2}{\lambda}&irJ_nJ_n'&r^2J_n^2
 \end{pmatrix}.
 \label{eq:harmonic-tensor}
\end{equation}
For real velocity variables and real $\mathbf{k}$, the integer-order Bessel functions and their derivatives are real. Thus $\mathsf{T}_{sn}$ has the structure
\begin{equation}
 \mathsf{T}_{sn}=\begin{pmatrix}
 A&iB&C\\
 -iB&D&-iE\\
 C&iE&F
 \end{pmatrix},
 \label{eq:cartesian-structure}
\end{equation}
where $A,\ldots,F$ are real functions. Complex conjugation changes the sign of each entry with exactly one $y$ index. Multiplication by $R$ on both sides changes precisely the same entries, giving
\begin{equation}
 R\mathsf{T}_{sn}^*R=\mathsf{T}_{sn}.
 \label{eq:harmonic-parity}
\end{equation}

The scalar factors have the corresponding conjugation property,
\begin{align}
 \Delta_{sn}(\omega^*)&=\Delta_{sn}(\omega)^*,\\
 U_s(\omega^*)&=U_s(\omega)^*.
 \label{eq:scalar-conjugation}
\end{align}
These identities show that every harmonic contribution satisfies
\begin{equation}
 \chi_{sn}(\omega^*,\mathbf{k})
 =R\chi_{sn}(\omega,\mathbf{k})^*R.
 \label{eq:harmonic-response-pt}
\end{equation}
The nonresonant contribution has the same property because $R\hat{\mathbf{z}}=\hat{\mathbf{z}}$ and its only frequency dependence is $1/\omega^2$. Velocity integration, summation over harmonics, and summation over species therefore give Eq.~\eqref{eq:response-theorem}. Once the reflection $R$ has been identified, the proof is straightforward. The important step is recognizing the PT symmetry of the standard kinetic response.

\subsection{Independent proof from the Qin--Phillips--Davidson representation}

The susceptibility can also be expressed without an infinite cyclotron-harmonic sum. Using the representation of Qin, Phillips, and Davidson~\cite{Qin2007}, we introduce
\begin{equation}
 a=\frac{\omega-k_\parallel v_\parallel}{\Omega_s},
 \label{eq:complex-order-variables}
\end{equation}
and the complex-order-Bessel combination
\begin{equation}
 Q(a,\lambda)=\frac{\pi a}{\sin\pi a}J_{-a}(\lambda)J_a(\lambda).
 \label{eq:bessel-product}
\end{equation}
The cyclotron resonance conditions $a=n$ are contained in the factor $\sin\pi a$. The species susceptibility takes the compact form
\begin{equation}
 \chi_s=\chi_s^{\mathrm{nr}}
 +\frac{\omega_{ps}^2}{\omega\Omega_s}
 \int d\Gamma\,v_\perp U_s\,\mathsf{T}(a,\lambda).
 \label{eq:compact-response}
\end{equation}
For the conventions of Eq.~\eqref{eq:harmonic-tensor}, the compact tensor can be written as
\begin{equation}
 \mathsf{T}(a,\lambda)=\begin{pmatrix}
 \mathcal{A}&i\mathcal{B}&\mathcal{C}\\
 -i\mathcal{B}&\mathcal{D}&-i\mathcal{E}\\
 \mathcal{C}&i\mathcal{E}&\mathcal{F}
 \end{pmatrix},
 \label{eq:compact-tensor}
\end{equation}
where, with $Q_{\lambda}=\partial Q/\partial \lambda$, its scalar entries are
\begin{align}
 \mathcal{A}&=\frac{a}{\lambda^2}(Q-1),&
 \mathcal{B}&=\frac{Q_{\lambda}}{2\lambda},\nonumber\\
 \mathcal{C}&=\frac{r}{\lambda}(Q-1),&
 \mathcal{E}&=\frac{rQ_{\lambda}}{2a},\nonumber\\
 \mathcal{D}&=\frac{\pi J_{-a}'(\lambda)J_a'(\lambda)}{\sin\pi a}
 +\frac{a}{\lambda^2},&
 \mathcal{F}&=\frac{r^2Q}{a}.
 \label{eq:compact-scalars}
\end{align}
Removable singularities in these expressions are evaluated by taking their limits.

For positive real $\lambda$, the Bessel functions obey $J_\nu(\lambda)^*=J_{\nu^*}(\lambda)$. Since $(\sin\pi a)^*=\sin\pi a^*$, it follows directly that
\begin{align}
 Q(a^*,\lambda)&=Q(a,\lambda)^*,\\
 Q_{\lambda}(a^*,\lambda)&=Q_{\lambda}(a,\lambda)^*,
 \label{eq:q-conjugation}\\
 \frac{\pi J_{-a^*}'(\lambda)J_{a^*}'(\lambda)}{\sin\pi a^*}
 &=\left[\frac{\pi J_{-a}'(\lambda)J_a'(\lambda)}{\sin\pi a}\right]^*.
 \label{eq:bessel-derivative-conjugation}
\end{align}
These identities also hold for negative signed $\lambda$ when the paired Bessel factors are continued consistently. Indeed, for $x>0$, the phases in $J_{-a}(-x)=e^{-i\pi a}J_{-a}(x)$ and $J_a(-x)=e^{i\pi a}J_a(x)$ cancel in their product. The derivative product has the same cancellation, while $Q_{\lambda}$ changes sign under $\lambda\to-\lambda$. Thus the scalar conjugation identities apply for either sign of $\Omega_s$.

Every scalar entry in Eq.~\eqref{eq:compact-scalars} consequently satisfies $\mathcal{A}(a^*,\lambda)=\mathcal{A}(a,\lambda)^*$, and likewise for $\mathcal{B},\ldots,\mathcal{F}$. Akin to the situation in Eq.~\eqref{eq:harmonic-tensor}, the explicit factors of $i$ occur only in entries with exactly one $y$ index. Therefore,
\begin{equation}
 \mathsf{T}(a^*,\lambda)=R\mathsf{T}(a,\lambda)^*R.
 \label{eq:compact-parity}
\end{equation}
Because $a(\omega^*)=a(\omega)^*$, Eq.~\eqref{eq:compact-response}, together with the scalar conjugation properties of $U_s$ and the nonresonant term, gives Eq.~\eqref{eq:response-theorem} after velocity integration and species summation.

This proof establishes the susceptibility symmetry directly in the compact representation, without resolving the response into separate cyclotron harmonics. The two proofs therefore identify the same PT symmetry in different representations of the kinetic response. This indicates a more fundamental symmetry of magnetized kinetic dynamics.

\subsection{Conductivity, dielectric tensor, and electromagnetic dispersion relation}

With the Fourier convention used in the Introduction, the induced current is $\mathbf{J}=\sigma\mathbf{E}$, and the conductivity and susceptibility are related in SI units by
\begin{equation}
 \sigma(\omega,\mathbf{k})=-i\epsilon_0\omega\chi(\omega,\mathbf{k}).
 \label{eq:conductivity-definition}
\end{equation}
Equation~\eqref{eq:response-theorem} then gives
\begin{equation}
 \sigma(\omega^*,\mathbf{k})=-R\sigma(\omega,\mathbf{k})^*R.
 \label{eq:conductivity-parity}
\end{equation}
For the relative dielectric tensor $\epsilon=I+\chi$, the corresponding relation is
\begin{equation}
 \epsilon(\omega^*,\mathbf{k})=R\epsilon(\omega,\mathbf{k})^*R.
 \label{eq:dielectric-parity}
\end{equation}

The electromagnetic dispersion matrix in the $\mathbf{E}$ frame obtained from Maxwell's equations is
\begin{equation}
 M(\omega,\mathbf{k})=\mathbf{k}\mathbf{k}-k^2I
 +\frac{\omega^2}{c^2}\epsilon(\omega,\mathbf{k}),
 \label{eq:dispersion-matrix-definition}
\end{equation}
where $\mathbf{k}\mathbf{k}$ is the dyadic product and $c$ is the speed of light. Since $R\mathbf{k}=\mathbf{k}$ and $R^2=I$, the vacuum Maxwell part satisfies
\begin{equation}
 R(\mathbf{k}\mathbf{k}-k^2I)R=\mathbf{k}\mathbf{k}-k^2I.
 \label{eq:vacuum-parity}
\end{equation}
Combining this identity with Eq.~\eqref{eq:dielectric-parity} proves Eq.~\eqref{eq:matrix-theorem}. Taking determinants and using $(\det R)^2=1$ yields Eq.~\eqref{eq:dispersion-pt}. Thus every zero of the formal electromagnetic dispersion relation has a complex-conjugate partner, completing the proof of Theorem~\ref{thm:susceptibility}.

\subsection{Why Does the Susceptibility Tensor Possess PT Symmetry?}
\label{sec:kinetic-origin}

The two preceding proofs establish PT symmetry of the standard hot-plasma susceptibility, but they do not identify its kinetic origin. The susceptibility is obtained only after the distribution-function perturbations have been eliminated. It is therefore natural to ask whether its PT symmetry is inherited from the underpinning kinetic system.

For a prescribed electromagnetic perturbation, the linearized Vlasov equation can be written schematically as
\begin{equation}
 \mathcal{L}_s(\omega,\mathbf{k})g_s=\mathcal{F}_s[\mathbf{E}],
 \label{eq:forced-kinetic-equation}
\end{equation}
where $\mathcal{L}_s$ contains streaming and gyro-rotation, and $\mathcal{F}_s$ contains the linearized force acting on the equilibrium distribution. Here the magnetic perturbation has been expressed in terms of $\mathbf{E}$ using Faraday's law. Wherever the inverse is defined, solving for the kinetic perturbation gives
\begin{equation}
 g_s=\mathcal{L}_s^{-1}(\omega,\mathbf{k})\mathcal{F}_s[\mathbf{E}].
 \label{eq:eliminated-kinetic-state}
\end{equation}
The induced current is then
\begin{equation}
 \mathbf{J}=\sum_s q_s\int\mathbf{v}g_s\,d^3v
 =\sigma(\omega,\mathbf{k})\mathbf{E}.
 \label{eq:reduced-current}
\end{equation}
Thus the response from $\mathbf{E}$ to $\mathbf{J}$ contains the effect of the eliminated infinite-dimensional kinetic dynamics. The susceptibility inherits this information through Eq.~\eqref{eq:conductivity-definition}.

The proofs of Theorem~\ref{thm:susceptibility} suggest that the response symmetry reflects a symmetry of the complete system before this elimination. We show in the next section that the Hamiltonian operator $\HVM$ of the linearized Vlasov--Maxwell system satisfies
\begin{equation}
 PT\HVM(PT)^{-1}=\HVM,
 \label{eq:underpinning-pt}
\end{equation}
where parity $P$ and time reversal $T=K$ are identified on the complete electromagnetic kinetic eigenmode space. The PT symmetry of $\chi$ is a reduced-response-space manifestation of this operator identity. Once the transformations of the fields and the kinetic perturbations are known, the susceptibility symmetry follows directly from the unreduced equations, without explicitly solving for $g_s$ or evaluating its velocity moments.

The susceptibility identity constrains the eigenfrequencies but does not specify parity on the kinetic variables. The unreduced eigenvector contains the distribution perturbations and both electromagnetic fields. Defining $P$ and $T$ on this complete state reveals whether a mode preserves PT symmetry and identifies its conjugate partner. We therefore return to the linearized Vlasov--Maxwell equations.

\section{PT Symmetry of the Linearized Vlasov--Maxwell System}
\label{sec:vlasov-maxwell}

We now identify the PT symmetry of the linearized Vlasov--Maxwell system. Parity and time reversal must therefore be defined on this full state space of the perturbed distribution functions and the electromagnetic fields. We first display the linearized Vlasov--Maxwell eigenproblem, identify the $P$ and $T$ operators, and then demonstrate invariance of the equations under their successive action. 

\subsection{Linearized Vlasov--Maxwell eigenproblem}
\label{sec:vm-eigenproblem}

The geometry and coordinate system are those specified in Secs.~\ref{sec:introduction} and \ref{sec:susceptibility}. For each species, write the distribution as
\begin{equation}
 f_s(\mathbf{x},\mathbf{v},t)=F_{0s}(\mathbf{v})
 +g_s(\mathbf{v})e^{i\mathbf{k}\cdot\mathbf{x}-i\omega t}.
 \label{eq:vm-distribution}
\end{equation}
The electric and magnetic perturbations have the same Fourier dependence, with amplitudes $\mathbf{E}$ and $\delta\mathbf{B}$. Here the equilibrium distribution function $F_{0s}$ is normalized to the number density, $\int F_{0s}\,d^3v=n_{0s}$. The wave vector is real and lies in the $x$--$z$ plane as in Eq.~\eqref{eq:geometry}. With $v_x=v_\perp\cos\theta$ and $v_y=v_\perp\sin\theta$, the linearized Vlasov equation for the perturbed distribution function $g_s$ is
\begin{align}
 \omega g_s={}&\left(k_\perp v_\perp\cos\theta
 +k_\parallel v_\parallel\right)g_s
 +i\Omega_s\frac{\partial g_s}{\partial\theta}\nonumber\\
 &-i\frac{q_s}{m_s}
 \left(\mathbf{E}+\mathbf{v}\times\delta\mathbf{B}\right)
 \cdot\nabla_vF_{0s}.
 \label{eq:vm-vlasov}
\end{align}

Faraday's and Ampere's laws give the dynamical field equations
\begin{align}
 \omega\delta\mathbf{B}&=\mathbf{k}\times\mathbf{E},
 \label{eq:vm-faraday}\\
 \omega\mathbf{E}&=-c^2\mathbf{k}\times\delta\mathbf{B}
 -\frac{i}{\epsilon_0}\mathbf{J},
 \label{eq:vm-ampere}
\end{align}
where the perturbed charge and current densities are
\begin{equation}
 \rho=\sum_s q_s\int g_s\,d^3v,\qquad
 \mathbf{J}=\sum_s q_s\int\mathbf{v}g_s\,d^3v.
 \label{eq:vm-moments}
\end{equation}
Physical perturbations also satisfy the Maxwell constraints $i\mathbf{k}\cdot\mathbf{E}=\rho/\epsilon_0$ and $\mathbf{k}\cdot\delta\mathbf{B}=0$.

Define the complete eigenvector by
\begin{equation}
 \Psi=\left(\{g_s\},\mathbf{E},\delta\mathbf{B}\right)^T.
 \label{eq:vm-state}
\end{equation}
Equations~\eqref{eq:vm-vlasov}--\eqref{eq:vm-ampere} then define the Hamiltonian operator $\HVM$ through
\begin{equation}
\omega\Psi= \HVM\Psi.
 \label{eq:vm-full-eigenproblem}
\end{equation}
The perturbed distribution functions are periodic in gyrophase and have well-defined velocity moments. We define $P$ and $T$ on the unconstrained amplitude space. Their product $PT$ preserves the Maxwell constraints and therefore acts within the physical eigenmode space.

\subsection{Gyrotropy follows from the homogeneous stationary equilibrium}
\label{sec:vm-gyrotropy}

Gyrotropy follows directly from the equilibrium Vlasov equation. For the homogeneous stationary equilibrium considered here, that equation reduces to 
\begin{equation}
 \frac{q_s}{m_s}(\mathbf{v}\times\mathbf{B}_0)
 \cdot\nabla_vF_{0s}
 =-\Omega_s\frac{\partial F_{0s}}{\partial\theta}=0.
 \label{eq:vm-equilibrium}
\end{equation}
For charged species in a nonzero magnetic field, this requires
\begin{equation}
 F_{0s}=F_{0s}(v_\perp,v_\parallel).
 \label{eq:vm-gyrotropic-equilibrium}
\end{equation}
Thus gyrotropy is a consequence of stationarity and homogeneity, rather than an additional symmetry assumption. Gyrotropy makes the equilibrium invariant under reflection across the $\mathbf{k}$--$\mathbf{B}_0$ plane. Differentiating this invariance and using $R^TR=I$ give
\begin{equation}
 F_{0s}(R\mathbf{v})=F_{0s}(\mathbf{v}),\qquad
 \nabla_vF_{0s}(R\mathbf{v})=R\nabla_vF_{0s}(\mathbf{v}).
 \label{eq:vm-equilibrium-reflection}
\end{equation}
No reflection symmetry in $v_\parallel$ is required.

\subsection{Parity transformation and time-reversal transformation}
\label{sec:vm-transformations}

We first define the parity operator $P$.  The distribution perturbation $g_s$ is a scalar function of velocity, $\rho$ is a scalar charge-density amplitude, and $\mathbf{E}$, $\delta\mathbf{B}$, and $\mathbf{J}$ are vector amplitudes. We define the action of $P$ separately on these quantities. 

For a scalar function of velocity, introduce the operator $\mathcal{R}_\theta$ that replaces $(v_x,v_y,v_\parallel)$ by $(v_x,-v_y,v_\parallel)$, or equivalently $\theta$ by $-\theta$:
\begin{equation}
 (\mathcal{R}_\theta g_s)(v_\perp,v_\parallel,\theta)
 =g_s(v_\perp,v_\parallel,-\theta).
 \label{eq:vm-gyrophase-reflection}
\end{equation}
We define $Pg_s=-\mathcal{R}_\theta g_s$ for the distribution perturbation, $P\mathbf{E}=R\mathbf{E}$ for the electric field, and $P\delta\mathbf{B}=-R\delta\mathbf{B}$ for the magnetic perturbation. For the charge and current densities, we define $P\rho=-\rho$ and $P\mathbf{J}=-R\mathbf{J}$. These definitions specify different actions on different scalar and vector amplitudes. On the complete eigenmode space, they give
\begin{equation}
 P\Psi=\left(\{-\mathcal{R}_\theta g_s\},
 R\mathbf{E},-R\delta\mathbf{B}\right)^T.
 \label{eq:vm-parity}
\end{equation}
These definitions are consistent with the kinetic moments in Eq.~\eqref{eq:vm-moments}. The change of variable $\mathbf{v}\mapsto R\mathbf{v}$ preserves the integration measure $d^3v$ because $|\det R|=1$. Substituting $Pg_s=-g_s(R\mathbf{v})$ into the charge moment gives $P\rho=-\rho$, and the current moment gives
\begin{equation}
 P\mathbf{J}=-\sum_s q_s\int\mathbf{v}g_s(R\mathbf{v})\,d^3v
 =-R\mathbf{J}.
 \label{eq:vm-parity-current}
\end{equation}
Both $R$ and $\mathcal{R}_\theta$ square to the identity. Consequently, $P$ is a linear involution satisfying $P^2=I$.

Time reversal is defined independently as complex conjugation of the normal-mode amplitudes,
\begin{equation}
 T=K,\qquad
 T\Psi=\left(\{g_s^*\},\mathbf{E}^*,\delta\mathbf{B}^*\right)^T.
 \label{eq:vm-time-reversal}
\end{equation}
This transformation is antilinear and satisfies $T^2=I$ and $TiT^{-1}=-i$. It is the time-reversal operation in the fixed-background spectral problem. Its relation to physical time reversal of the nonlinear Vlasov--Maxwell equations will be discussed in Sec.~\ref{sec:physical-time-reversal}. Neither $P$ nor $T$ is required to commute with $\HVM$ . We next examine invariance under their successive action.

\subsection{PT symmetry of the VM system}
\label{sec:vm-pt-proof}

We now verify PT symmetry by substituting the defined scalar and vector transformations into the coupled equations. Applying $P$ after complex conjugation gives
\begin{equation}
 PT\Psi=\left(\{-\mathcal{R}_\theta g_s^*\},
 R\mathbf{E}^*,-R\delta\mathbf{B}^*\right)^T.
 \label{eq:vm-combined-transformation}
\end{equation}
Denote the components of this transformed state by $\widetilde g_s$, $\widetilde{\mathbf{E}}$, and $\widetilde{\delta\mathbf{B}}$. The velocity argument is reflected only in $\widetilde g_s$. The vector amplitudes are transformed at the same wave vector, and the charge and current amplitudes satisfy $\widetilde\rho=-\rho^*$ and $\widetilde{\mathbf{J}}=-R\mathbf{J}^*$. We first verify the kinetic equation and then the Maxwell equations. Throughout, $\mathbf{k}$ and $\mathbf{B}_0$ remain fixed.

The streaming factor $k_\perp v_\perp\cos\theta+k_\parallel v_\parallel$ is real and even in $\theta$. It is therefore invariant under both complex conjugation and gyrophase reflection. For gyro-rotation,
\begin{equation}
 \mathcal{R}_\theta\partial_\theta\mathcal{R}_\theta^{-1}
 =-\partial_\theta,
 \label{eq:vm-derivative-reflection}
\end{equation}
so the sign change of $\partial_\theta$ compensates the sign change of $i$ under time reversal. Writing the free kinetic operator as
\begin{equation}
 \mathcal{A}_s=k_\perp v_\perp\cos\theta
 +k_\parallel v_\parallel+i\Omega_s\partial_\theta,
 \label{eq:vm-free-operator}
\end{equation}
these properties give $\mathcal{R}_\theta\mathcal{A}_s^*\mathcal{R}_\theta^{-1}=\mathcal{A}_s$.

For the electric-force term, substituting $\widetilde{\mathbf{E}}=R\mathbf{E}^*$ and using Eq.~\eqref{eq:vm-equilibrium-reflection} gives
\begin{equation}
 \widetilde{\mathbf{E}}\cdot\nabla_vF_{0s}(\mathbf{v})
 =\mathbf{E}^*\cdot\nabla_vF_{0s}(R\mathbf{v}).
 \label{eq:vm-electric-force-reflection}
\end{equation}
For the magnetic-force term, substitute $\widetilde{\delta\mathbf{B}}=-R\delta\mathbf{B}^*$. The vector identity $(R\mathbf{a})\times(R\mathbf{b})=(\det R)R(\mathbf{a}\times\mathbf{b})$, together with $R^2=I$, gives
\begin{equation}
 \mathbf{v}\times(-R\delta\mathbf{B}^*)
 =R\left[(R\mathbf{v})\times\delta\mathbf{B}^*\right]
 \label{eq:vm-magnetic-cross-product}
\end{equation}
and the same equilibrium-gradient relation implies
\begin{align}
 &\left(\mathbf{v}\times\widetilde{\delta\mathbf{B}}\right)
 \cdot\nabla_vF_{0s}(\mathbf{v})\nonumber\\
 &\qquad=\left[(R\mathbf{v})\times\delta\mathbf{B}^*\right]
 \cdot\nabla_vF_{0s}(R\mathbf{v}).
 \label{eq:vm-magnetic-force-reflection}
\end{align}
Thus each transformed force term equals its complex conjugate evaluated at the reflected velocity. Taking the complex conjugate of Eq.~\eqref{eq:vm-vlasov} and applying $-\mathcal{R}_\theta$ therefore yields
\begin{align}
 \omega^*\widetilde g_s={}&\mathcal{A}_s\widetilde g_s
 -i\frac{q_s}{m_s}
 \left(\widetilde{\mathbf{E}}
 +\mathbf{v}\times\widetilde{\delta\mathbf{B}}\right)
 \cdot\nabla_vF_{0s}.
 \label{eq:vm-transformed-vlasov}
\end{align}
This is the original kinetic equation at $(\omega^*, PT\Psi)$. 

We next verify the Maxwell equations. Since $R\mathbf{k}=\mathbf{k}$, the same cross-product identity gives
\begin{equation}
 \mathbf{k}\times R\mathbf{E}^*
 =-R(\mathbf{k}\times\mathbf{E}^*).
 \label{eq:vm-faraday-reflection}
\end{equation}
Complex conjugation of Faraday's law followed by multiplication by $-R$ gives
\begin{equation}
 \omega^*\widetilde{\delta\mathbf{B}}
 =\mathbf{k}\times\widetilde{\mathbf{E}}.
 \label{eq:vm-transformed-faraday}
\end{equation}
For Ampere's law, use $\widetilde{\mathbf{J}}=-R\mathbf{J}^*$. Complex conjugation followed by multiplication by $R$ gives
\begin{align}
 \omega^*\widetilde{\mathbf{E}}
 &=R\left[-c^2\mathbf{k}\times\delta\mathbf{B}^*
 +\frac{i}{\epsilon_0}\mathbf{J}^*\right]\nonumber\\
 &=-c^2\mathbf{k}\times\widetilde{\delta\mathbf{B}}
 -\frac{i}{\epsilon_0}\widetilde{\mathbf{J}}.
 \label{eq:vm-transformed-ampere}
\end{align}
Finally, $\widetilde\rho=-\rho^*$, complex conjugation of Gauss's law, and $R\mathbf{k}=\mathbf{k}$ give $i\mathbf{k}\cdot\widetilde{\mathbf{E}}=\widetilde\rho/\epsilon_0$. The magnetic constraint is also preserved, since $\mathbf{k}\cdot\widetilde{\delta\mathbf{B}}=-\mathbf{k}\cdot\delta\mathbf{B}^*=0$. Thus the defined transformations preserve the complete coupled eigenproblem under $PT$.

\begin{theorem}
\label{thm:vlasov-maxwell}
\textup{(PT symmetry of the linearized Vlasov--Maxwell system)}
For a homogeneous collisionless magnetized plasma with a stationary gyrotropic equilibrium, let $\HVM$ be the operator defined in Eq.~\eqref{eq:vm-full-eigenproblem}. The parity and time-reversal transformations defined in Eqs.~\eqref{eq:vm-parity} and \eqref{eq:vm-time-reversal} satisfy
\begin{equation}
 PT\HVM(PT)^{-1}=\HVM.
 \label{eq:vm-operator-theorem}
\end{equation}
If $\HVM\Psi=\omega\Psi$, then
\begin{equation}
 \HVM(PT\Psi)=\omega^*(PT\Psi).
 \label{eq:vm-mode-pairing}
\end{equation}
The formal eigenfrequencies therefore occur in conjugate pairs,
\begin{equation}
 \omega\in\sigma_p(\HVM)\quad\Longrightarrow\quad
 \omega^*\in\sigma_p(\HVM).
 \label{eq:vm-spectrum-pairing}
\end{equation}
\end{theorem}

Here, $\sigma_p(\HVM)$ denotes the point spectrum in the chosen PT-invariant domain.  The invariance established above proves Eq.~\eqref{eq:vm-operator-theorem}. Acting with $PT$ on the eigenvalue equation and using its antilinearity gives Eq.~\eqref{eq:vm-mode-pairing}, which establishes the conjugate spectral structure. 

\subsection{Susceptibility symmetry from the unreduced kinetic equations}
\label{sec:vm-response-reduction}

Theorem~\ref{thm:vlasov-maxwell} gives the susceptibility symmetry without an explicit solution of the linearized Vlasov equation. To obtain the response, take $\mathbf{E}$ as a prescribed input to the kinetic equations and express $\delta\mathbf{B}$ through Faraday's law. The self-consistent Ampere equation is imposed after this reduction. Wherever the kinetic response is defined, the field produces a current $\mathbf{J}=\sigma(\omega,\mathbf{k})\mathbf{E}$. PT symmetry maps this response to the field $R\mathbf{E}^*$ and current $-R\mathbf{J}^*$ at frequency $\omega^*$. The same response relation therefore requires
\begin{equation}
 \sigma(\omega^*,\mathbf{k})R\mathbf{E}^*
 =-R\sigma(\omega,\mathbf{k})^*\mathbf{E}^*.
 \label{eq:vm-transformed-response}
\end{equation}
Since the relation holds for arbitrary field amplitudes, it follows that
\begin{equation}
 \sigma(\omega^*,\mathbf{k})=-R\sigma(\omega,\mathbf{k})^*R.
 \label{eq:vm-conductivity-result}
\end{equation}
Using $\sigma=-i\epsilon_0\omega\chi$ then gives
\begin{equation}
 \chi(\omega^*,\mathbf{k})=R\chi(\omega,\mathbf{k})^*R.
 \label{eq:vm-susceptibility-result}
\end{equation}
Thus the kinetic equations determine the susceptibility PT symmetry without an explicit calculation of the perturbed distributions or the response tensor. The two calculations in Sec.~\ref{sec:susceptibility} provide independent direct verification of this operator result. The PT symmetry of the familiar hot-plasma response is inherited from the PT symmetry of the complete electromagnetic kinetic system. 

\section{PT Symmetry of the Magnetized Vlasov--Poisson System}
\label{sec:vlasov-poisson}

\subsection{Linearized Vlasov--Poisson eigenproblem}

We now consider the Vlasov-Poisson system, which can be viewed as an approximate special case of the Vlasov-Maxwell system. Therefore, its exact PT symmetry needs to be independently established.  The electrostatic approximation retains the kinetic distributions and the self-consistent scalar potential. We set $\delta\mathbf{B}=0$ and $\mathbf{E}=-i\mathbf{k}\phi$, and determine $\phi$ from Poisson's equation. This defines the Vlasov--Poisson system, with the dynamical Maxwell equations replaced by the electrostatic field equation. For $k\ne0$, the potential is the linear functional
\begin{equation}
 \Phi[g]=\frac{1}{\epsilon_0k^2}\sum_{s}q_{s}\int g_{s}\,d^3v.
 \label{eq:vp-potential}
\end{equation}
The complete kinetic state is $g=\{g_s\}$, and its eigenvalue equation is
\begin{align}
\omega g=&\HVP g,\\
 (\HVP g)_s={}&\left(k_\perp v_\perp\cos\theta
 +k_\parallel v_\parallel\right)g_s
 +i\Omega_s\partial_\theta g_s\nonumber\\
 &-\frac{q_s}{m_s}\left(k_\perp\cos\theta
 \frac{\partial F_{0s}}{\partial v_\perp}
 +k_\parallel\frac{\partial F_{0s}}{\partial v_\parallel}\right)\Phi[g].
 \label{eq:vp-operator}
\end{align}
The electrostatic coupling follows from substituting $\mathbf{E}=-i\mathbf{k}\phi$ into Eq.~\eqref{eq:vm-vlasov}. Equation~\eqref{eq:vp-operator} retains the gyrophase and velocity dependence of every species and is the unreduced electrostatic eigenproblem.

\subsection{Parity and time reversal in the Vlasov--Poisson system}

On the electrostatic kinetic state, define parity by gyrophase reflection,
\begin{equation}
 (Pg)_s(v_\perp,v_\parallel,\theta)
 =g_s(v_\perp,v_\parallel,-\theta).
 \label{eq:vp-parity}
\end{equation}
This is a linear involution with $P^2=I$ and $P\partial_\theta P^{-1}=-\partial_\theta$. The definition of time reversal is unchanged, i.e., $T=K$. Their successive action compensates for the two sign changes in the gyro-rotation term $i\Omega_s\partial_\theta$.

The streaming term is real and even in $\theta$. The coefficient multiplying $\Phi[g]$ in Eq.~\eqref{eq:vp-operator} has the same properties because the equilibrium is real and gyrotropic. Changing the gyrophase integration variable gives
\begin{equation}
 \Phi[Pg]=\Phi[g],\qquad \Phi[Tg]=\Phi[g]^*.
 \label{eq:vp-potential-transform}
\end{equation}
Consequently, the self-consistent electrostatic coupling is also invariant under $PT$. Gyrophase periodicity is preserved by both transformations.

The kinetic parity in Eq.~\eqref{eq:vp-parity} differs by an overall minus sign from the kinetic component of Eq.~\eqref{eq:vm-parity}. This is a choice of representation on the electrostatic state. Multiplying the full electromagnetic parity by $-1$ leaves its operator symmetry unchanged and gives the electrostatic transformation $g\mapsto\mathcal{R}_\theta g^*$, $\phi\mapsto\phi^*$. Thus the electrostatic and electromagnetic representations are consistent.

\begin{theorem}
\label{thm:vlasov-poisson}
\textup{(PT symmetry of the magnetized Vlasov--Poisson eigenproblem)}
For a homogeneous collisionless magnetized plasma with a stationary gyrotropic equilibrium, let $\HVP$ be the operator defined in Eq.~\eqref{eq:vp-operator}. The transformations $P=\mathcal{R}_\theta$ and $T=K$ satisfy
\begin{equation}
 PT\HVP(PT)^{-1}=\HVP.
 \label{eq:vp-theorem}
\end{equation}
If $\HVP g=\omega g$, then
\begin{equation}
 \HVP(PTg)=\omega^*(PTg).
 \label{eq:vp-pairing}
\end{equation}
The eigenfrequencies therefore occur in conjugate pairs, and the electrostatic dispersion function obeys
\begin{equation}
 D_{\mathrm{VP}}(\omega^*,\mathbf{k})
 =D_{\mathrm{VP}}(\omega,\mathbf{k})^*.
 \label{eq:vp-dispersion-symmetry}
\end{equation}
\end{theorem}

The term-by-term invariance established above proves Eq.~\eqref{eq:vp-theorem}. Antilinearity then gives Eq.~\eqref{eq:vp-pairing}. To establish the dispersion relation statement, prescribe $\phi$ in the kinetic equation and write the induced charge as $\rho=\mathcal{Q}(\omega,\mathbf{k})\phi$. The transformed solution has potential $\phi^*$ and charge $\rho^*$ at frequency $\omega^*$, so $\mathcal{Q}(\omega^*,\mathbf{k})=\mathcal{Q}(\omega,\mathbf{k})^*$. Poisson's equation gives
\begin{equation}
 D_{\mathrm{VP}}(\omega,\mathbf{k})
 =1-\frac{\mathcal{Q}(\omega,\mathbf{k})}{\epsilon_0k^2},
 \label{eq:vp-response-dispersion}
\end{equation}
which proves Eq.~\eqref{eq:vp-dispersion-symmetry} without explicitly solving for the perturbed distribution. The conjugation symmetry of the electrostatic dispersion relation is therefore inherited from PT symmetry of the unreduced kinetic equations, akin to the situation in the Vlasov-Maxwell system.

\section{Physical Consequences and Predictive Power of PT Symmetry in Magnetized Kinetic Plasmas}
\label{sec:physical-consequences}

The PT symmetry established in Secs.~\ref{sec:vlasov-maxwell} and \ref{sec:vlasov-poisson} constrains kinetic instability. A growing eigenmode with frequency $\omega_r+i\gamma$ must have a decaying partner with frequency $\omega_r-i\gamma$. The kinetic operator preserves PT symmetry, while the individual unstable eigenmodes break it spontaneously.

In this section, we connect this spontaneous PT breaking to Krein collision \cite{Krein1950,Yakubovich75,Zhang2016GH}.  PT symmetry entails pseudo-Hermiticity, and the transition from real frequencies to a complex-conjugate pair requires a collision of modes with opposite action signatures. At a generic two-mode collision, the dispersion function has a real double root, and the frequency splitting follows a square-root law in the distance from threshold. This structure supplies an analytical criterion for instability onset and is particularly relevant to nonmonotonic distributions and to resonant energetic-particle populations.

Magnetic mirrors in fusion energy research provide an important physical setting. Post and Rosenbluth analyzed electrostatic loss-cone instabilities in finite mirror-confined plasmas.\cite{PostRosenbluth1966} Subsequent work examined finite-beta stabilization,\cite{Tang1972DCLC} the dependence of drift-cyclotron loss-cone (DCLC) instability on radial density gradients,\cite{Ferron1984DCLC} and DCLC dynamics in simulations of tandem mirrors with sloshing ions.\cite{Cohen1984DCLC} These topics are being actively pursued in contemporary devices \cite{Forest2024,Kolmes2024LossCone,Frank2025,Tran2025DCLC,Shih2026}. These studies motivate examining the spectral structure of velocity-space-driven instabilities. We present three cases in which PT symmetry has distinct utilities: it explains thermal Bernstein stability, yields an analytical description of DGH onset, and guides the discovery of mode-pole instability.

\subsection{PT-symmetry breaking and Krein collision}
\label{sec:krein-collision}

As established in Sec.~\ref{sec:pt-overview}, Eqs.~\eqref{eq:partner} and \eqref{eq:preserving} require a real frequency for a PT-preserving eigenmode. Spontaneous PT breaking instead gives the growing and decaying conjugate pair in Eqs.~\eqref{eq:broken} and \eqref{eq:growing-decaying}, while the operator retains the symmetry in Eq.~\eqref{eq:general-pt}. We now examine how the transition occurs through a process called Krein collision.

\begin{figure*}[tb]
 \centering
 \includegraphics[width=0.49\textwidth]{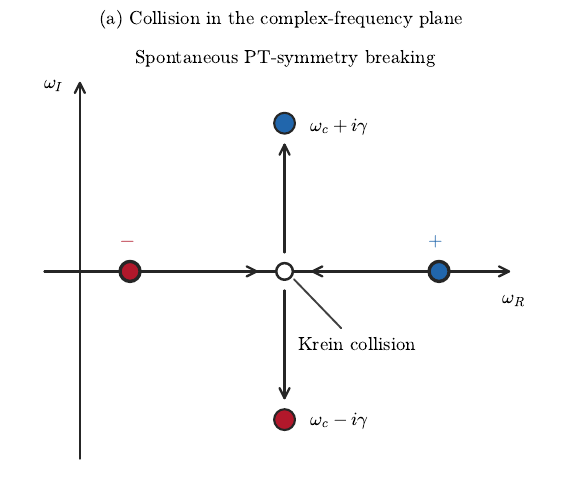}\hfill
 \includegraphics[width=0.49\textwidth]{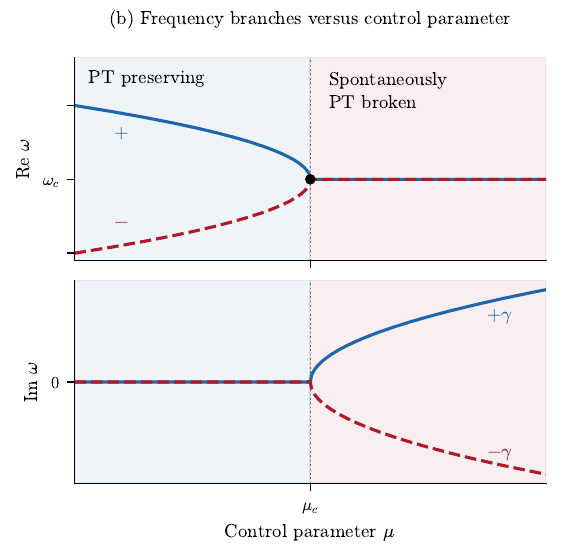}
 \caption{Schematic Krein collision and spontaneous PT-symmetry breaking. (a) Trajectories in the complex-frequency plane. Real modes with opposite action signatures, indicated by $+$ and $-$, approach the collision at $\omega_c$ and emerge as a growing and decaying conjugate pair. Arrows indicate the direction of increasing control parameter, and the open circle marks the collision. (b) The corresponding frequency branches versus $\mu$. Two real branches meet at $\mu_c$; beyond the collision their real parts coincide and their imaginary parts are $+\gamma$ and $-\gamma$. The curves illustrate the local square-root form in Eq.~\eqref{eq:collision-splitting}; all axes have arbitrary scales.}
 \label{fig:krein-collision}
\end{figure*}

Let a real parameter $\mu$ vary continuously in a dispersion relation $D(\omega,\mu)=0$. Away from its singularities, a generic collision of two real roots satisfies
\begin{equation}
 D(\omega_c,\mu_c)=0,\qquad
 D_\omega(\omega_c,\mu_c)=0.
 \label{eq:general-double-root}
\end{equation}
Here subscript $\omega$ denotes $\partial/\partial\omega$. If $D_{\omega\omega}\ne0$ and $D_\mu\ne0$ at the collision, expansion about this point gives
\begin{equation}
 0=D_\mu\delta\mu+\tfrac12D_{\omega\omega}\delta\omega^2
 +\cdots.
 \label{eq:collision-expansion}
\end{equation}
The leading frequency splitting is consequently
\begin{equation}
 \delta\omega_\pm\sim\pm
 \sqrt{-\frac{2D_\mu}{D_{\omega\omega}}\delta\mu}.
 \label{eq:collision-splitting}
\end{equation}
On one side of the transition the two frequencies are real. On the other side they form a complex-conjugate pair, as illustrated in Fig.~\ref{fig:krein-collision}.

For a pseudo-Hermitian system, such a transition is called Krein collision \cite{Krein1950,Yakubovich75,Zhang2016GH}. A system is pseudo-Hermitian if its Hamiltonian  $\mathcal{H}$ satisfies $\mathcal{H}^\dagger G=G\mathcal{H}$ for a nonsingular Hermitian metric $G$. A simple real-frequency mode carries the signature of $\Psi^\dagger G\Psi$, the sign of the corresponding mode action. Definite-sign subspaces remain spectrally stable under sufficiently small perturbations that preserve this structure. The creation of a complex pair therefore requires collision of modes with opposite signatures.\cite{Krein1950,Yakubovich75,Zhang2016GH} Opposite signatures permit instability; a collision does not by itself guarantee that the roots leave the real axis.

We proved that finite-dimensional PT symmetry entails pseudo-Hermiticity regardless of diagonalizability.\cite{Zhang2020PT} This connects the PT symmetry breaking to the Krein collision. At the neighborhood of  a collision, the system can be approximated by a two-dimsional system. The only route from real frequencies to a complex pair under a structure-preserving parameter variation is a Krein collision involving opposite signatures.

\subsection{Thermal Bernstein modes: protection by positive cyclotron weights}
\label{sec:bernstein-stability}

For strictly perpendicular propagation, $k_\parallel=0$, the thermal electrostatic Bernstein dispersion relation has the meromorphic form\cite{Bernstein1958}
\begin{equation}
 D_{\mathrm B}(z)=C-\sum_n\frac{A_n}{z-z_n},
 \qquad A_n>0,\quad z_n>0,
 \label{eq:bernstein-meromorphic}
\end{equation}
where $z=\omega^2$ and $C>0$ is independent of frequency. For a single kinetic species, $z_n=n^2\Omega^2$, $n\ge1$, and $C=1$. Multiple thermal species contribute additional positive weights at their cyclotron poles; coincident poles can be combined. We call $A_n$ the cyclotron weights. In the sign convention of Eq.~\eqref{eq:bernstein-meromorphic}, the actual complex-analysis residue is $-A_n$.

On the real axis away from the poles,
\begin{equation}
 D_{\mathrm B}'(z)=\sum_n\frac{A_n}{(z-z_n)^2}>0.
 \label{eq:bernstein-monotonicity}
\end{equation}
The dispersion function increases from $-\infty$ to $+\infty$ between successive distinct poles and has exactly one simple root in each interval. 

We infer that complex eigenfrequencies do not exist because the positivity of the cyclotron weights $A_n$ prevents the degeneracy required for a real-mode collision. The Bernstein branches remain simple and interlace with the poles, and spontaneous PT symmetry breaking cannot occur within this thermal perpendicular electrostatic problem. This identifies the response property that protects the known Bernstein stability.

\subsection{Dory--Guest--Harris instability: analytical PT-breaking mechanism}
\label{sec:dgh-instability}

The DGH equilibrium distribution is gyrotropic but nonmonotonic in perpendicular velocity and can support kinetic instabilities.\cite{Dory1965,Kolmes2024LossCone,Qin2026IBW,Chen2026} We apply the Krein collision analysis of Sec.~\ref{sec:krein-collision} to identify its spontaneous PT-symmetry breaking mechanism. For perpendicular propagation, the dispersion function retains the structure of Eq.~\eqref{eq:bernstein-meromorphic}, with poles $z_n=n^2\Omega^2$, but the cyclotron weights $A_n$ need not be positive. A parameter $\mu$ may control the density, perpendicular wave number, or distribution shape.

For a finite-frequency collision away from the poles, the double-root condition in Eq.~\eqref{eq:general-double-root} becomes $D(z_c,\mu_c)=D_z(z_c,\mu_c)=0$, with $z_c>0$. The derivative in Eq.~\eqref{eq:bernstein-monotonicity} can vanish only if the nonzero weights include both signs. Thus the DGH distribution must remove the monotonicity that protects the thermal Bernstein branches before a collision is possible. Opposite signs alone do not determine the threshold; the dispersion relation and its derivative must vanish at the same point. The square-root splitting follows from Eq.~\eqref{eq:collision-splitting} with $\omega$ replaced by $z$, and the corresponding frequency shift is $(z-z_c)/(2\sqrt{z_c})$. The distribution-dependent weights and the resulting threshold are calculated below.

To calculate the weights explicitly, consider one kinetic species with density $n_0$ and an immobile neutralizing background. Write $F_0=n_0f_0$ and choose the normalized DGH distribution \cite{Dory1965}
\begin{equation}
 f_0^{(j)}=
 \frac{x^j e^{-x}}{\pi v_t^2 j!}
 \frac{e^{-v_\parallel^2/v_{\parallel t}^2}}
 {\sqrt\pi v_{\parallel t}},
 \qquad x=\frac{v_\perp^2}{v_t^2},
 \label{eq:dgh-equilibrium}
\end{equation}
where $j$ is a nonnegative integer. Its perpendicular distribution  is $h_j(v_\perp)=\int f_0^{(j)}\,dv_\parallel$, with $2\pi\int_0^\infty v_\perp h_j\,dv_\perp=1$. The case $j=0$ is thermal, while $j>0$ has a maximum at $v_r=\sqrt j\,v_t$ and a depleted low-perpendicular-velocity population.

For a general normalized perpendicular distribution $h$, pairing the positive and negative harmonics of the electrostatic response gives
\begin{equation}
 A_n=-\frac{4\pi\omega_p^2 n^2\Omega^2}{k_\perp^2}
 \int_0^\infty\frac{dh}{dv_\perp}
 J_n^2\!\left(\frac{k_\perp v_\perp}{|\Omega|}\right)dv_\perp.
 \label{eq:general-cyclotron-weight}
\end{equation}

Substituting Eq.~\eqref{eq:dgh-equilibrium}, and defining $\kappa=k_\perp v_t/|\Omega|$, yields the weights
\begin{align}
 A_n^{(j)}={}&\frac{4\omega_p^2 n^2}{\kappa^2 j!}
 \int_0^\infty e^{-x}\left(x^j-jx^{j-1}\right)
 J_n^2(\kappa\sqrt x)\,dx.
 \label{eq:dgh-exact-weight}
\end{align}
For $j>0$, the factor $x^j-jx^{j-1}$ changes sign at the maximum of the distribution, and different harmonics can acquire opposite weights.

The narrow-ring limit makes the Bessel coupling especially explicit. Substitution of $h(v_\perp)=\delta(v_\perp-v_r)/(2\pi v_r)$ into Eq.~\eqref{eq:general-cyclotron-weight}, followed by integration by parts, gives
\begin{equation}
 A_n^{\mathrm{ring}}=
 \frac{2\omega_p^2n^2}{\lambda}\frac{dJ_n^2(\lambda)}{d\lambda},
 \qquad \lambda=\frac{k_\perp v_r}{|\Omega|}.
 \label{eq:ring-exact-weight}
\end{equation}
The derivative changes sign with the ring radius sampled by each harmonic. 

For a numerical verification, hold $j$ and $\kappa$ fixed and vary $\mu=\omega_p^2/\Omega^2$. With $\zeta=z/\Omega^2$ and $a_n=A_n^{(j)}/\omega_p^2$, the dispersion relation is
\begin{equation}
 D(\zeta,\mu)=1-\mu S(\zeta),\qquad
 S(\zeta)=\sum_{n\ge1}\frac{a_n}{\zeta-n^2}.
 \label{eq:dgh-dimensionless}
\end{equation}
The analytical threshold is determined by
\begin{equation}
 \sum_{n\ge1}\frac{a_n}{(\zeta_c-n^2)^2}=0,
 \qquad \mu_c=\frac{1}{S(\zeta_c)}>0.
 \label{eq:dgh-threshold-evaluation}
\end{equation}
For $j=8$ and $\kappa=4.5/\sqrt8$, the first four weights are approximately $0.0104$, $-0.1353$, $-0.1457$, and $0.3098$. The two branches between the first and second cyclotron poles collide at
\begin{align}
 \zeta_c&=1.6523,  \mu_c=25.8465,
 \frac{\omega_c}{|\Omega|}=1.2854.
 \label{eq:dgh-numerical-threshold}
\end{align}
At this point $D_{\zeta\zeta}=-2.4946$ and $D_\mu=-1/\mu_c$. Increasing $\mu$ through $\mu_c$ therefore produces a complex pair, with
\begin{equation}
 \frac{\gamma}{|\Omega|}\sim
 \frac{1}{2\sqrt{\zeta_c}}
 \sqrt{-\frac{2}{D_{\zeta\zeta}}
 \left(\frac{\mu}{\mu_c}-1\right)}.
 \label{eq:dgh-dimensionless-growth}
\end{equation}

\begin{figure*}[tb]
 \centering
 \includegraphics[width=0.49\textwidth]{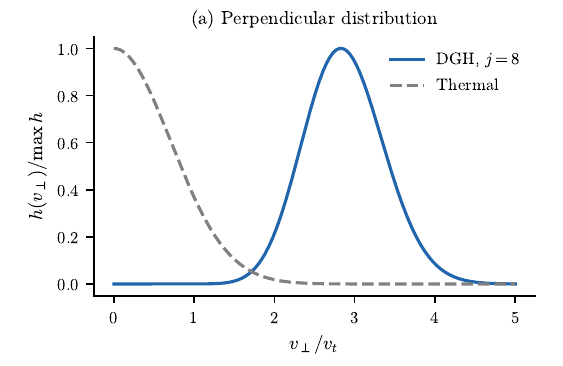}\hfill
 \includegraphics[width=0.49\textwidth]{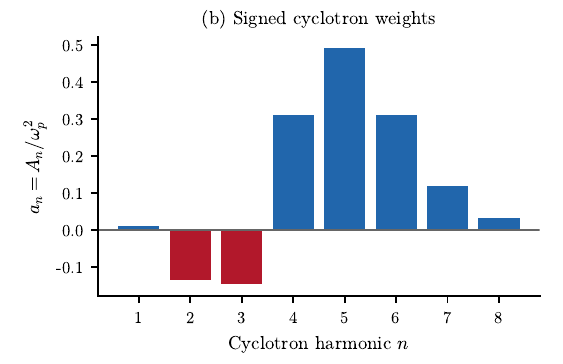}\\[1ex]
 \includegraphics[width=0.49\textwidth]{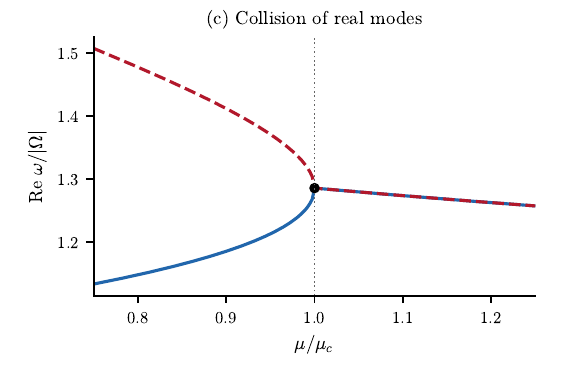}\hfill
 \includegraphics[width=0.49\textwidth]{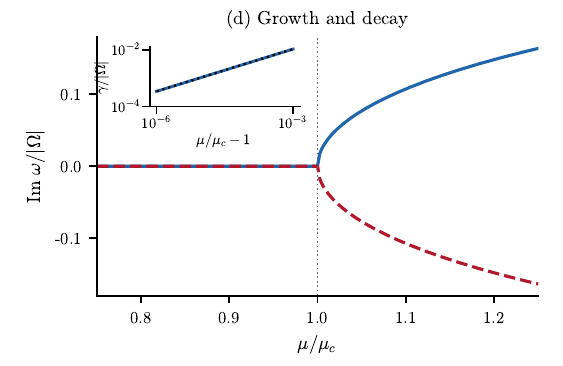}
 \caption{Spontaneous PT breaking for the smooth DGH equilibrium in Eq.~\eqref{eq:dgh-equilibrium}, with $j=8$, $k_\perp v_t/|\Omega|=4.5/\sqrt8$, and an immobile neutralizing background. Panel (a) shows the perpendicular distribution and a thermal distribution, each normalized to its own maximum. Panel (b) shows the first eight signed cyclotron weights. Panel (c) follows the real parts of the two positive-frequency branches as $\mu=\omega_p^2/\Omega^2$ crosses $\mu_c=25.8465$. Panel (d) shows the conjugate growth and decay rates. Its inset compares the computed growth rate (solid) with the analytical square-root law (dotted).}
 \label{fig:dgh-transition}
\end{figure*}

Figure~\ref{fig:dgh-transition} tracks these roots through the transition. The calculation uses 40 harmonics and 128-point generalized Gauss--Laguerre quadrature. Changing these to 20 harmonics and 64 points, or 60 harmonics and 192 points, changes $\mu_c$ by less than $10^{-8}$. The tracked roots satisfy $|D|<10^{-10}$. A logarithmic fit over $10^{-6}\le\mu/\mu_c-1\le10^{-3}$ gives a growth-rate exponent of $0.49998$, consistent with Eq.~\eqref{eq:dgh-dimensionless-growth}.

In this example, the PT symmetry analysis identifies the triggering mechanism of this DGH instability. The nonmonotonic equilibrium distribution produces sign-indefinite cyclotron weights, which remove the monotonicity protecting thermal Bernstein modes. The double-root condition then determines when two real modes collide and become an unstable-damped pair.

\subsection{Mode-pole-resonance instability: PT symmetry as a guide to discovery}
\label{sec:mode-pole-instability}

PT-symmetry analysis suggests a distinct collision mechanism when a collective plasma mode approaches a pole introduced by an energetic minority population. When the additional population is weak, one root of the full dispersion relation lies close to the background eigenfrequency $\omega_0$, while another lies close to the energetic-particle pole $\omega_p$. These are two approximate eigenfrequencies of the coupled system. Bringing them close together therefore identifies a candidate Krein collision and, for opposite action signatures, a possible instability through spontaneous PT symmetry breaking. This observation directs the search for instability toward mode-pole resonance.

Suppose the background dispersion has a simple real root, $D_0(\omega_0)=0$, and the additional response has a nearby simple pole at $\omega_p$. Locally,
\begin{equation}
 D_0(\omega)\simeq a(\omega-\omega_0),\qquad
 \chi_p(\omega)\simeq\frac{b}{\omega-\omega_p},
 \label{eq:mode-pole-expansion}
\end{equation}
where $a=D_0'(\omega_0)\ne0$, and $a$, $b$, and $\omega_p$ are real. In this subsection $\omega_p$ denotes the pole frequency, rather than the plasma frequency used above. The local dispersion relation becomes
\begin{equation}
 a(\omega-\omega_0)+\frac{b}{\omega-\omega_p}=0.
 \label{eq:mode-pole-dispersion}
\end{equation}
For small nonzero $b$ and separated frequencies, $|b/a|\ll|\omega_0-\omega_p|^2$, the two roots approach $\omega_0$ and $\omega_p$.  For nonzero $b$, multiplication by $\omega-\omega_p$ gives a quadratic with roots
\begin{equation}
 \omega_\pm=\frac{\omega_0+\omega_p}{2}
 \pm\sqrt{\left(\frac{\omega_0-\omega_p}{2}\right)^2-\frac{b}{a}}.
 \label{eq:mode-pole-roots}
\end{equation}
For $b/a>0$, reducing the detuning brings the two real roots together. They undergo a Krein collision when the squared half-detuning equals $b/a$, and the inequality
\begin{equation}
 \frac{b}{a}>\left(\frac{\omega_0-\omega_p}{2}\right)^2
 \label{eq:mode-pole-condition}
\end{equation}
gives a pair of complex-conjugate eigenfrequencies.  At exact resonance, $\omega_p=\omega_0$, any positive $b/a$ yields
\begin{equation}
 \omega_\pm=\omega_0\pm i\sqrt{\frac{b}{a}}.
 \label{eq:mode-pole-exact-resonance}
\end{equation}
If $b$ is proportional to the energetic population's density, the growth rate scales as the square root of that density and has no finite density threshold at exact resonance. At finite coupling the two roots have the Krein collision structure shown in Fig.~\ref{fig:krein-collision}.

The sign in Eq.~\eqref{eq:mode-pole-condition} has a direct physical interpretation. Near a positive minority cyclotron pole, $\omega_p=n|\Omega_h|$, the ring response in Eq.~\eqref{eq:ring-exact-weight} has
\begin{equation}
 b=-\frac{A_n^{\mathrm{ring}}}{2\omega_p}.
 \label{eq:mode-pole-ring-residue}
\end{equation}
For a thermal background Bernstein root at positive frequency, $a>0$ follows from Eq.~\eqref{eq:bernstein-monotonicity}. A negative ring weight therefore supplies the sign needed for instability. The minority pole must be distinct from the background cyclotron poles, so that $D_0$ is regular near the resonant collective mode. As shown in Ref.~\onlinecite{Qin2026IBW}, the negative ring weight represents the free energy associated with the nonmonotonic distribution of the energetic ions. 

This mode-pole mechanism was identified in the study of thresholdless ion Bernstein wave emission and alpha-to-thermal-ion energy channeling. \cite{Qin2026IBW} The alpha cyclotron pole resonates with a thermal proton Bernstein mode and produces a complex conjugate pair of modes that channels the fusion energy directly to the thermal fuel ions. Alpha channeling  \cite{FischRax1992,Ochs2021,Ochs2022,fisch1995alpha2,FischHerrmann1999Tutorial,Fisch2006Mirror,FettermanFisch2008Rotating,ZhmoginovFisch2009Waves} could be  an enabling technology for $p$--$^{11}\mathrm{B}$ fusion \cite{Rostoker1993,Rostoker1997,Putvinski2019,Kolmes2022a,Ochs2022,wei2023,Kong2023,Liu2024,Ochs2024,Kamio2026} and other aneutronic advanced fuel fusion \cite{Qin2024AFF}. PT-symmetry theory guided the discovery by directing attention to the resonance and the sign of the mode-pole coupling. 

\section{Conclusion and Discussion}
\label{sec:discussion}

\subsection{Conclusions}
\label{sec:conclusion}

We have shown that the hot-plasma susceptibility satisfies $\chi(\omega^*,\mathbf{k})=R\chi(\omega,\mathbf{k})^*R$ for a homogeneous stationary gyrotropic equilibrium, where $R=\operatorname{diag}(1,-1,1)$ reflects velocity across the plane containing $\mathbf{k}$ and $\mathbf{B}_0$. Direct calculations in the cyclotron-harmonic and compact susceptibility representations establish the same identity.

This response symmetry originates in the unreduced kinetic equations. By defining parity on the distribution perturbations and electromagnetic amplitudes and combining it with complex conjugation, we proved that the linearized Vlasov--Maxwell operator commutes with $PT$. An independent construction establishes the corresponding result for Vlasov--Poisson. The susceptibility symmetry survives elimination of the kinetic variables, while the full operator symmetry specifies the conjugate partner of each eigenmode.

Many kinetic instabilities driven by nonmonotonic distributions, rings, loss cones, and resonant energetic-particle populations can be understood as spontaneous PT-symmetry breaking. Krein collisions between modes of opposite action provide the only route to instability in these PT-symmetric systems.

Three examples are given. Thermal Bernstein waves illustrate stability maintained by positive cyclotron weights. For the DGH equilibrium, the same analysis yields a Krein collision threshold and a square-root growth law, verified by numerical continuation. Resonance between a collective mode and a kinetic pole illustrates its predictive use in identifying fusion-product-driven ion Bernstein wave emission. These results provide a framework for explaining, deriving, and discovering kinetic instabilities in magnetized plasmas. 

\subsection{Nonlinear time-reversal symmetry vs PT symmetry}
\label{sec:physical-time-reversal}

Interpreting these results requires distinguishing spectral time reversal from nonlinear time reversal. The full collisionless nonlinear Vlasov--Maxwell equations possess physical time-reversal symmetry. If $f_s$, $\mathbf{E}$, and $\mathbf{B}$ constitute a solution, then another solution is given by
\begin{align}
 f_s^{\mathrm{tr}}(\mathbf{x},\mathbf{v},t)
 &=f_s(\mathbf{x},-\mathbf{v},-t),\nonumber\\
 \mathbf{E}^{\mathrm{tr}}(\mathbf{x},t)&=\mathbf{E}(\mathbf{x},-t),\nonumber\\
 \mathbf{B}^{\mathrm{tr}}(\mathbf{x},t)&=-\mathbf{B}(\mathbf{x},-t).
 \label{eq:nonlinear-time-reversal}
\end{align}
Physical time reversal acts on the full time-dependent solution. For plasma waves in a magnetized equilibrium plasma, it reverses all velocity components and maps the equilibrium to one with magnetic field $-\mathbf{B}_0$ and distribution $F_{0s}(-\mathbf{v})$. It therefore relates dynamics about two backgrounds that need not coincide.

The linear PT symmetry established in Secs.~\ref{sec:vlasov-maxwell} and \ref{sec:vlasov-poisson} instead acts on complex mode amplitudes within a fixed equilibrium and at fixed real $\mathbf{k}$. Its velocity reflection changes $v_y$ but leaves $v_\parallel$ unchanged, and its field transformations do not reverse the prescribed $\mathbf{B}_0$. The transformation pairs eigenfrequencies $\omega$ and $\omega^*$ of the same kinetic operator.

The spectral operation $T=K$ must therefore be distinguished from the full nonlinear map in Eq.~\eqref{eq:nonlinear-time-reversal}. At fixed equilibrium, the spectral constraint follows from the combined action of $P$ and $T$ defined earlier.

\subsection{Formal kinetic eigenproblem vs Landau pole of causal response}
\label{sec:landau-discussion}

A further distinction concerns the formal kinetic eigenproblem and the causal response of an initial-value problem. Theorems~\ref{thm:vlasov-maxwell} and \ref{thm:vlasov-poisson} concern the formal collisionless eigenproblems $\HVM\Psi=\omega\Psi$ and $\HVP g=\omega g$. At fixed real $\mathbf{k}$, their operator equations and admissibility conditions are preserved by $PT$. An eigenmode in a PT-invariant domain therefore has its conjugate partner in the same domain. This is the operator statement underlying the response identities derived using real velocity integrals. It must be distinguished from a choice of causal continuation for an initial-value problem.\cite{Stix1992}

For $k_\parallel\ne0$, the harmonic response contains the denominator
\begin{equation}
 \omega-k_\parallel v_\parallel-n\Omega_s,
 \qquad
 v_{\parallel,\mathrm{res}}=\frac{\omega-n\Omega_s}{k_\parallel}.
 \label{eq:parallel-resonance-discussion}
\end{equation}
For nonreal $\omega$, the resonance lies away from the real velocity contour. Conjugating $\omega$ reflects this pole across the real $v_\parallel$ axis. Whenever the integrals converge, evaluating the response directly over real velocities on either side gives the conjugation relation used in the formal eigenproblem. On the real frequency axis, resonant denominators require additional care, and the kinetic operator can have a continuous spectrum.\cite{VanKampen1955,Case1959}

The causal response by Landau is constructed differently.\cite{Landau1946} A Fourier--Laplace transform first defines it in an upper half-plane where the time transform converges; for an unstable plasma the initial inversion contour lies above the growing poles. When the equilibrium permits analytic continuation, the response is continued toward the real axis and into the lower half-plane using the Landau prescription, with the velocity contour deformed as a resonance crosses it. The continued expression therefore need not equal the lower-half-plane integral evaluated directly on the undeformed real velocity axis. The difference contains the resonant contribution selected by the causal prescription.

These two spectral objects must consequently be distinguished. A formal eigenvalue belongs to the unreduced operator with its specified domain.  A Landau pole is a pole of the analytically continued causal field response and need not correspond to a regular decaying eigenfunction in the original state space. The PT theorem directly constrains the first object. A growing eigenmode and its formal PT-conjugate decaying partner need not appear as conjugate poles on the same causal analytic sheet. In particular, Landau damping does not require a growing partner on that sheet.

\section{Future Work}
\label{sec:future-work}
The operator symmetry established here motivates further work on kinetic wave topology and on explicit PT symmetry breaking mechanisms for instabilities beyond the collisionless model. 

\subsection{Topology of PT-symmetric kinetic systems}
\label{sec:kinetic-topology}

Topological plasma physics is an emerging research field that uses the geometry and topology of plasma eigenmodes to identify and study wave phenomena\cite{Parker2020a,Parker2020,Parker2021,Fu2021,Qin2023,Fu2022,MehrpourBernety2023,MehrpourBernety2024,Rajawat2025,Xie2025,Rao2025,Billings2026WTLCW}. Related developments concern photon and relativistic plasma-wave topology \cite{Palmerduca2024,Qin2024,Palmerduca2025}. 

Currently, most topological plasma physics is limited to Hermitian systems. The Vlasov--Maxwell and Vlasov--Poisson operators studied here are generally non-Hermitian in their usual kinetic representations, but they possess PT symmetry. Their topological physics can therefore involve real spectra, complex-conjugate branches, and transitions between these regimes. Our initial study established a relation between bulk topology and spectral flow in inhomogeneous PT-symmetric continuous media with non-Hermitian interface operators, and identified a topological Alfv\'en sound wave in Hall magnetohydrodynamics.\cite{FuQin2024Topology} This demonstrates that PT symmetry can support topological wave phenomena beyond Hermitian systems. Extending this direction to the full kinetic state space is expected to reveal new modes and new relations between topology and stability.

A first task is to determine which isolated branches of $\HVM$ and $\HVP$ admit well-defined topological invariants and how these invariants change near spontaneous PT-symmetry breaking. Spatially varying equilibria would then allow us to investigate kinetic interface modes and spectral flow. The present homogeneous theory supplies a starting point for these questions.

These developments may have useful applications to fusion energy research. If kinetic topological waves retain robust propagation along an interface, engineered density and magnetic-field profiles might guide radio-frequency power toward regions where resonant particles absorb it. Their interaction with energetic ions could also offer a way to influence fusion-product transport or channel fusion-product energy into thermal ions, complementing the mode-pole mechanism discussed in Sec.~\ref{sec:physical-consequences}.\cite{Qin2026IBW}. Mirroring applications in photonics, topological waves in plasma can be utilized to ensure waves propagate and transfer momentum strictly in one direction. This avoids bi-directional current cancellation and significantly maximizes steady-state current drive efficiency \cite{fisch87}.

\subsection{Explicit PT-symmetry breaking}
\label{sec:extensions}

\begin{figure}[tb]
 \centering
 \includegraphics[width=\columnwidth]{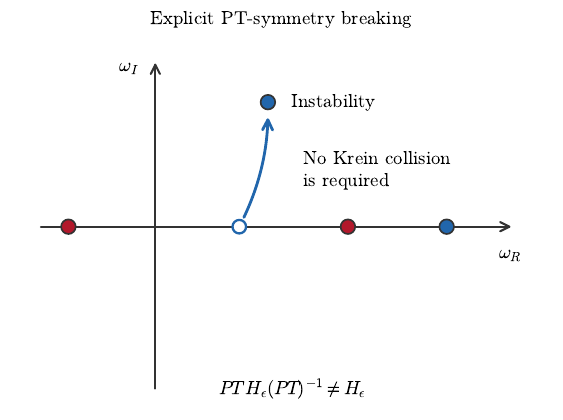}
 \caption{Schematic destabilization by explicit PT-symmetry breaking, following the concept illustrated in Refs.~\onlinecite{Qin2021,Qin2026TS}. The open circle marks the initial real eigenfrequency of the mode that becomes unstable. The arrow shows its motion into the upper half-plane as an explicit PT symmetry breaking perturbation of strength $\epsilon$ increases. Filled circles mark the frequencies of the perturbed operator $H_\epsilon$; the other three modes remain on the real axis. No Krein collision is required, and the frequencies need not form complex-conjugate pairs. Colors identify branches, and the axes have arbitrary scales.}
 \label{fig:explicit-pt-breaking}
\end{figure}

Alongside the study of topology within PT-symmetric systems, explicit PT-symmetry breaking provides another direction for kinetic plasma physics. As demonstrated in the present study, when the governing operator remains PT symmetric, a simple isolated real eigenfrequency cannot leave the real axis by itself under a regular variation of parameters. Spontaneous destabilization of the discrete branches considered here requires a Krein collision. If a physical process breaks the operator symmetry explicitly, however, the conjugation constraint is removed. An initially stable mode can then acquire a positive imaginary frequency without first colliding with another mode, as illustrated in Fig.~\ref{fig:explicit-pt-breaking}.

This mechanism was demonstrated for drift waves in magnetized plasmas using a fluid model.\cite{Qin2021} Electron--ion collisions break PT symmetry explicitly and allow equilibrium gradients to destabilize modes without the collision required in the conservative model. It was subsequently shown that viscosity can destabilize a two-stream system in parameter regimes that are stable in its absence.\cite{Qin2026TS} Dissipation in this example permits the release of free energy that is inaccessible to an instability while the PT constraint is intact. Explicit symmetry breaking does not guarantee growth, but it permits an instability route that the PT-symmetric operator forbids.

It will be very interesting to determine how this mechanism extends to the Vlasov--Maxwell and Vlasov--Poisson systems. A natural starting point is to add a physically specified collision operator to a collisionless PT-symmetric kinetic equilibrium and calculate the displacement of its isolated eigenfrequencies. It may identify collisional or dissipative boundary-induced instabilities that are absent from collisionless stability diagrams, and help determine how weak dissipation changes drift-wave transport and energetic-particle confinement. It may also suggest controlled ways of accessing energetic-particle free energy for useful wave excitation and heating.

\begin{acknowledgments}
This work was supported by  NSF Award PHY-2640873 (formerly PHY-2605414), Topological Waves in Non-Hermitian Plasmas and Enabling Technologies, and by flexible research funds from the University of Wisconsin--Madison, including funds provided by the Wisconsin Alumni Research Foundation (WARF). I thank Carl Bender, Nathaniel Fisch, Eric Palmerduca, Yichen Fu, and Asher Yahalom for fruitful discussions. This paper is dedicated to the memory of Prof. Ronald C. Davidson and Dr. Cynthia K. Phillips.
\end{acknowledgments}

\bibliographystyle{aipnum4-2}
\bibliography{references}
\end{document}